\documentclass[12pt,a4paper,dvipdfm]{article}
\usepackage{graphicx} 
\usepackage{comment}
\usepackage{setspace} 

\date{\empty}

\title{
\begin{flushleft}
{\large \bf
Transmission ERDA of ubiquitous deuterium in H-containing materials
}
\end{flushleft}
}
\begin{document}
\maketitle
\vspace{-15mm}
\author{\noindent 
Hiroshi Kudo$^{1}$
\footnote{Corresponding author\, {\it E-mail address}: kudo@tac.tsukuba.ac.jp} 
, Hiroshi Naramoto$^{1}$, Masao Sataka$^{1}$, Satoshi Ishii$^{1}$, Kimikazu Sasa$^{1, 2}$, Shigeo Tomita$^{2}$
}
\\\\  
$^{1}${\it CRiES, University of Tsukuba, 1-1-1 Tennodai, Tsukuba, Ibaraki 305-8577, Japan}\\
$^{2}${\it Institute of Pure and Applied Sciences, University of Tsukuba, 1-1-1 Tennodai, Tsukuba, Ibaraki 305-8571, Japan}\\\\
ABSTRACT\\
\setstretch{1.5}

\indent 
Transmission-type elastic recoil detection analysis  (T-ERDA) using 8.02 MeV He has been successfully 
applied to determine the concentration of ubiquitous deuterium in mylar and PPS films, 
both containing hydrogen as a component. The determined D/H ratios of $(1.60\pm 0.08) \times 10^{-4}$ and 
$(1.23\pm 0.08) \times 10^{-4}$ for mylar and PPS, respectively,  are only  
slightly deviated from the Standard Mean Ocean Water value of $1.55\cdots \times 10^{-4}$. 
These results indicate potential sensitivity of T-ERDA for analysis of the variation  
of ubiquitous deuterium in hydrogen-containing materials.  Such deuterium analysis might be   
useful for unique characterization of not only natural, but also synthesized hydrogen-containing materials.    

\newpage

\section{\large Introduction}\label{intro}

The hydrogen isotope D (deuterium, $^2$H), mixed in H (protium, $^1$H), has been of 
particular significance in astrophysics \cite{Rob2006}\cite{Gen2008} and 
geoscience \cite{Lec1998}\cite{Leh2022},  because the ratio of D to H content in 
matter (so-called D/H ratio) provides information about the formation process 
of the Solar System. The D/H ratio is changeable, for example, it increases if there are natural 
or artificial processes at work that concentrate it, such as distillation (because of higher boiling point of  
D-water than that of H-water) or electrolysis of water \cite{Lew1933}. Change in the D/H ratio is expected 
to occur in H-containing materials chemically synthesized on the earth, in which ubiquitous, i.e., 
naturally mixed, D is contained inherently. The D/H ratio seems to be a very similar fraction on the order of 
$10^{-4}$. The similar ratio is expected to be observed, for example, for synthetic resin 
containing H as a component. It is likely that 
analysis of ubiquitous D can be used to characterize synthesized materials. 

The recently developed T-ERDA for H is probably used for analysis of ubiquitous D 
in a thin solid sample if the background signals can be reduced enough to identify D signals  
\cite{Kud2021}\cite{Kud2022}, The major advantage of T-ERDA is that for both H and D 
the recoil cross sections at a recoil angle of $\sim$0$^{\circ}$ are enhanced to as large as 
1--2\,b/sr for projectile $^{4}$He energy of 8--10\,MeV \cite{iba}\cite{Gal1955}\cite{Ish2004}, 
which allows detection of trace amount of D and, accordingly, determination of the D/H ratio.
Furthermore, T-ERDA requires no reference sample in the analysis procedure \cite{Kud2022}. 
This feature is in contrast to the reported case of coincidence ERDA applied to 
the analysis of H and D in meteorite and mineral samples \cite{Ros2015}. 
On the basis of these characteristic aspects of T-ERDA, the purpose of the present 
study is to investigate the applicability of T-ERDA to the trace amount of D, i.e., on the 
order of D/H $\sim10^{-4}$, in an H-containing material.   

\section{ \large Experiments }\label{exp}

The samples used in the present experiments are thin H-containing films of mylar (polyethylene 
terephthalate, (C$_{10}$H$_{8}$O$_{4})_{n}$) of 2.5 $\mu$m thickness, and PPS (polyphenylene sulfide, 
(C$_{6}$H$_{4}$S)$_{n}$) of 1.35 $\mu$m thickness. These chemically synthesized products were 
chosen as typical samples containing hydrogen.  The PPS sample was cut from the same sheet as that 
previously used in the quantitative H analysis by T-ERDA \cite{Kud2022}. The mylar and PPS films 
are underlaid with a Ni foil of 20 or 25 $\mu$m thickness to prevent the incident He from entering 
the particle detector.  In this paper, the samples are labeled as mylar/Ni(20), PPS/Ni(20), 
and PPS/Ni(25), where the numbers in the parentheses denote the thicknesses in $\mu$m. 

\begin{figure}[h] 
\begin{center}
\includegraphics[width=150mm]{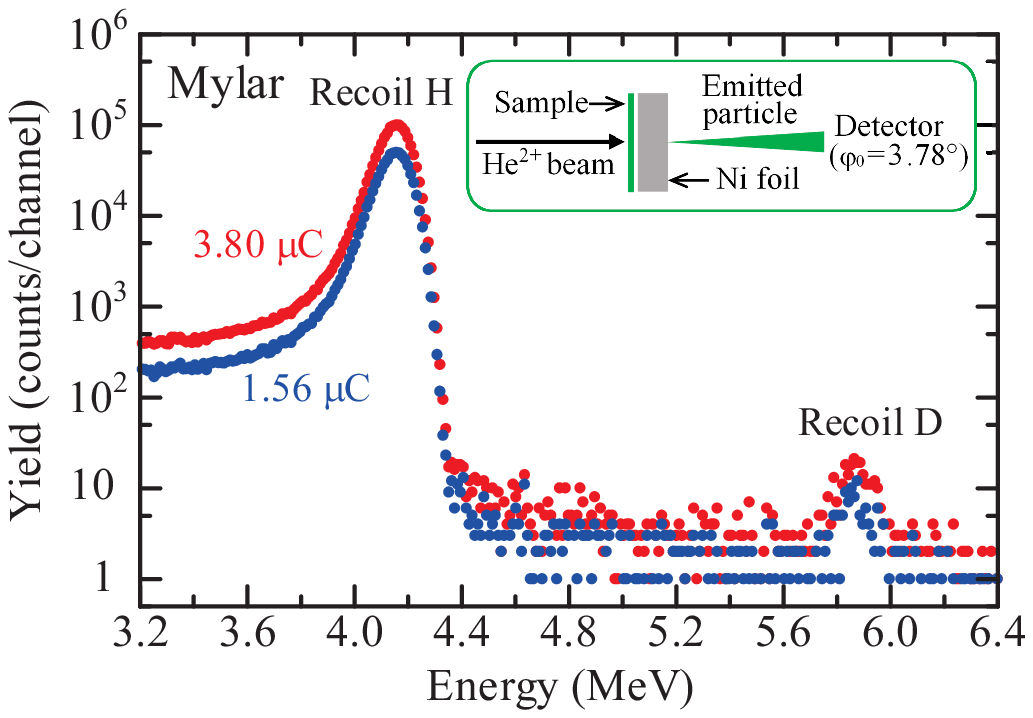}
\end{center}
\caption{T-ERDA spectra for incidence of 8.02 MeV He$^{2+}$ on the mylar film of 2.5 $\mu$m thickness, 
which were measured for the incident charges of 1.56 and 3.80 $\mu$C during the measurements. 
On the horizontal axis, 10.88 keV corresponds to 1\,channel width.
The inset schematically shows the experimental setup. }
\label{fig-1}
\end{figure}

The experimental setup is similar to that described previously \cite{Kud2022}. 
By using an incident beam of 8.02 MeV $^{4}$He$^{2+}$ from the UTTAC 6MV tandem accelerator, the experiments 
were carried out under the pressure of $2 \times 10^{-5}$ Pa.  The beam was focused typically to 
$\sim$0.1$\times$0.1\,mm$^{2}$ on the samples. The energy-measuring system was 
calibrated with 5.48 MeV $\alpha$ particles emitted from $^{241}$Am. 
The energies of recoil H and D were measured after passing through the Ni foil mainly by the surface-barrier 
silicon detector of 33 keV resolution for $\sim$5 MeV $\alpha$ particles, which accepts 
an angular range of 0 to $\varphi_{0} = 3.78^{\circ}$, see the inset of Fig. \ref{fig-1}. 
In a few cases, a pin-photodiode detector was successfully used for collecting the data.  
It is noted that the estimated angular spread of the incident He$^{2+}$ beam is $\sim$0.2$^{\circ}$, 
which is much less than $\varphi_{0}$.

The beam current was frequently monitored during the measurements by the Faraday cup attached to 
the sample holder.  The integrated beam current (incident charge) for 15 minutes was measured repeatedly 
with a current integrator, from which the typical uncertainty of the incident charge $\pm 7$\% was obtained.
The T-ERDA spectra were accumulated for 15--60 min for the He$^{2+}$ beam current of 
0.5--0.7 nA. Sometimes, short-time measurements for fresh beam spots on the sample  
were repeated two or three times and,  after confirming the reproducibility, these spectra were summed 
up to obtain the final data. This is because the H content in mylar and PPS decreases appreciably by the 
He$^{2+}$ irradiation under the present experimental conditions, which will be noted in \S\ref{DH-yield-ratio}. 

\section{ \large Results and discussion }\label{discuss}
\subsection{\normalsize Ratio of D to H yield}  
\label{DH-yield-ratio}

\begin{figure}[h] 
\begin{center}
\includegraphics[width=140mm]{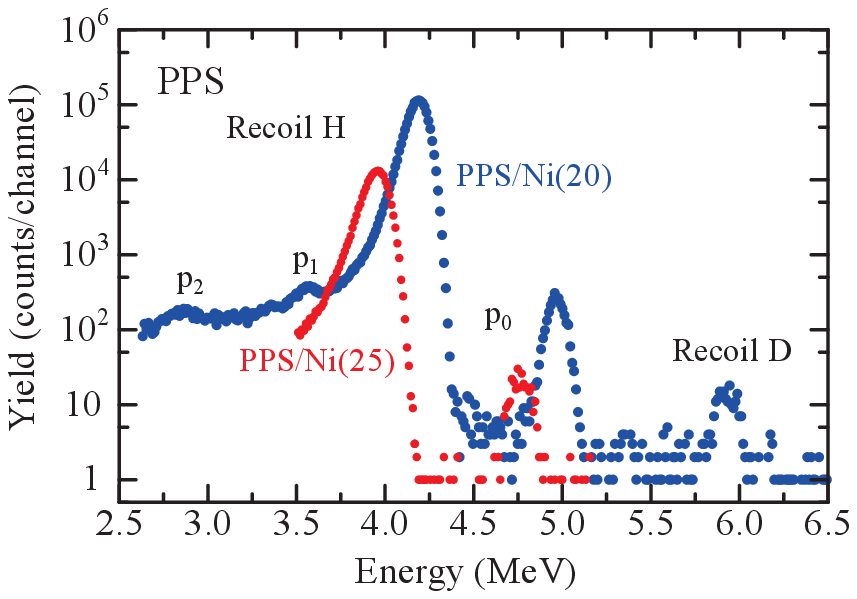}
\end{center}
\caption{T-ERDA spectra for incidence of 8.02 MeV He$^{2+}$ on the PPS film of 1.35 $\mu$m thickness, 
which were measured for the underlying Ni foils of 20 and 25 $\mu$m thicknesses.  The incident charges 
during the measurements are 4.00 and 0.90 $\mu$C for PPS/Ni(20) and PPS/Ni(25), respectively.  
On the horizontal axis, 10.97 keV corresponds to 1\,channel width.
Note that the p$_{0}$, p$_{1}$, and p$_{2}$ peaks are discussed later in \ref{S-reaction}. }
\label{fig-2}
\end{figure}

Figure \ref{fig-1} shows T-ERDA spectra of the mylar/Ni(20) sample, measured for the incident 
He$^{2+}$ charges of 1.56 and 3.80 $\mu$C. The two spectra are shown to confirm the 
appearance of a small peak at 5.87 MeV. From the estimate based on the stopping powers 
by SRIM \cite{Zie2013}, the two peaks at 4.16 and 5.87 MeV are identified as recoil H and D 
from the mylar film, respectively. 
Figure \ref{fig-2} shows T-ERDA spectra of the PPS/Ni(20) and PPS/Ni(25) samples, measured for 
the incident He$^{2+}$ charges of 4.00 and 0.90 $\mu$C, respectively.  
For PPS/Ni(20), the two peaks at 4.19 and 5.93 MeV are due to recoil H and D, respectively, 
from the PPS film, according to the similar estimate as for mylar/Ni(20). 
It is noteworthy that in Figs. 1 and 2 the D peaks have been successfully observed on the suppressed 
background yield by adopting the Ni foil of higher Coulomb barrier than that of the Al foil used previously.
The prominent peak at 4.97 MeV and relatively small peaks at 3.6 and 2.9 MeV will be discussed in 
\ref{S-reaction} by referring to the spectrum of PPS/Ni(25), in which the observed peak energies 
shift to the low-energy side because of the increased thickness of Ni by 5 $\mu$m. 

\begin{figure}[h] 
\begin{center}
\includegraphics[width=150mm]{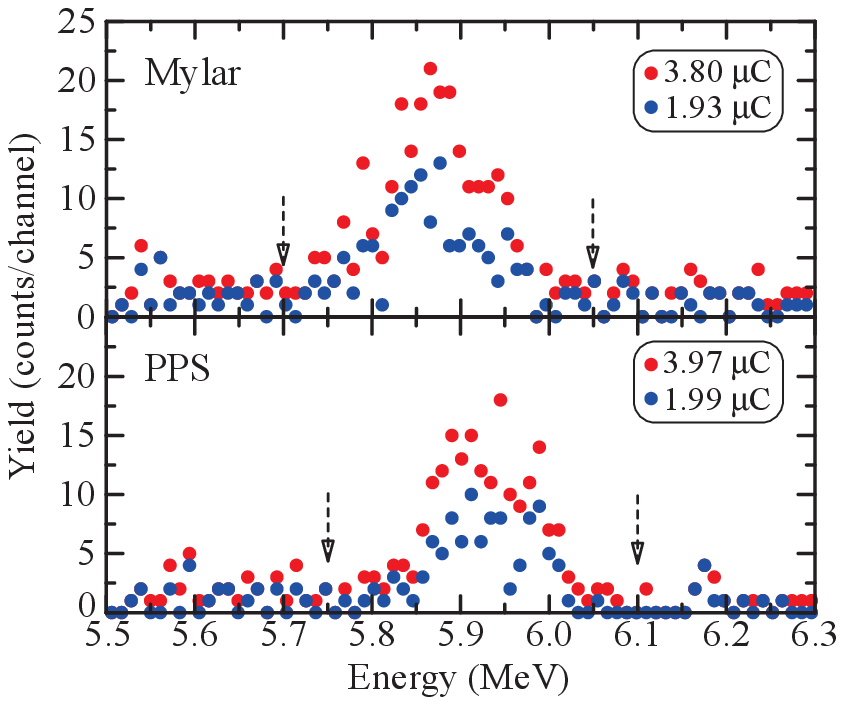}
\end{center}
\caption{Typical energy spectra of recoil D from the mylar/Ni(20) and PPS/Ni(20) samples, 
each measured  for the two incident charges of 8.02 MeV He$^{2+}$. The dashed arrows indicate 
the assumed low- and high-energy ends of the D peaks. }
\label{fig-3}
\end{figure}

From the T-ERDA spectra, D and H yields are obtained from the areas under the peaks after 
subtracting the discernible background yield in Figs. 1 and 2. The background subtraction has been 
carried out by using the straight baseline connecting the low- and high-energy background levels of each  
peak, which were obtained by averaging the yield over the $\pm$0.05 MeV range of the assumed low- and 
high-energy ends of the peak. Note that such ends of the peak are clearly recognized 
in every spectrum plotted in the linear scale, as shown for the D peaks in Fig. \ref{fig-3}.  

\begin{figure}[h] 
\begin{center}
\includegraphics[width=150mm]{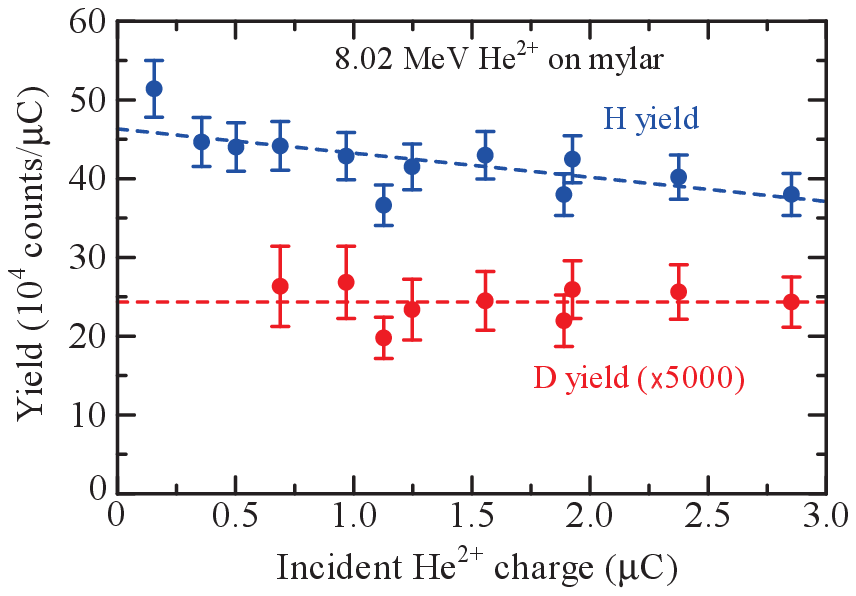}
\end{center}
\caption{H and D yield plotted against the incident He$^{2+}$ charge on mylar during the measurements. 
The dashed line for the H yield shows the least-square fit, while that for the D yield 
shows the averaged value. The value of $R=1.04 \times 10^{-4}$ for mylar is given by
the ratio of D to H yield at zero incident He$^{2+}$ charge on the dashed lines. }
\label{fig-4}
\end{figure}

\begin{figure}[h] 
\begin{center}
\includegraphics[width=150mm]{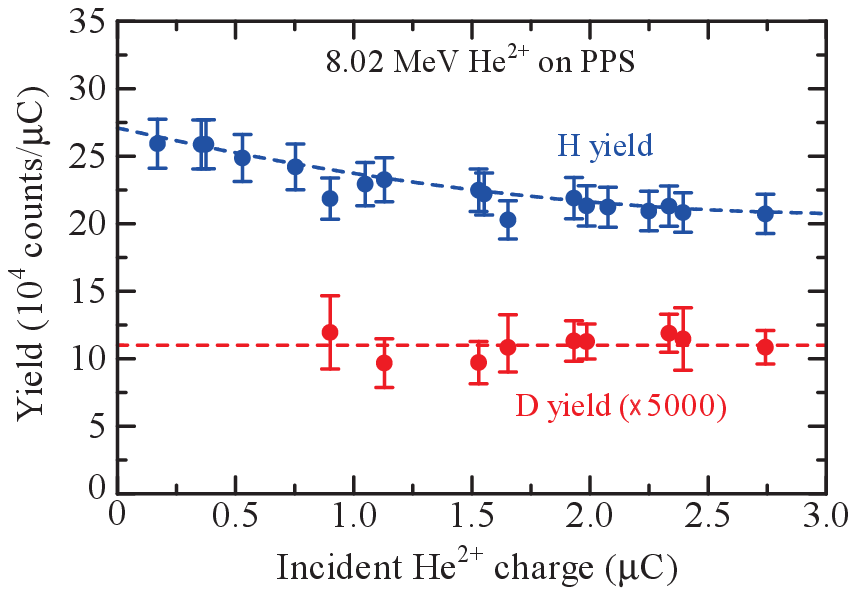}
\end{center}
\caption{H and D yield measured for PPS, presented similar to Fig. \ref{fig-4}. 
The value of $R=0.82 \times 10^{-4}$ for PPS is given by
the ratio of D to H yield at zero incident He$^{2+}$ charge on the dashed lines.}
\label{fig-5}
\end{figure}

Figures \ref{fig-4} and \ref{fig-5} show dependence of the H and D yields on the incident He$^{2+}$ charge on 
mylar and PPS, respectively. The vertical axes in counts/$\mu$C represent the yields normalized to 
the incident charge, which were obtained from the measured yields typically on the order of 10$^{2}$ 
and 10$^{6}$ counts for D and H, respectively. The error bars of the D yield are predominantly 
due to the uncertainty arising from the limited counts of the yield (statistical uncertainty), while those of the 
H yield are mainly due to the stability of the incident He$^{2+}$ charge ($\pm 7$\% noted in \S\ref{exp}). 

In Figs. \ref{fig-4} and \ref{fig-5}, we see appreciable decrease of the H yield with increasing 
the incident He$^{2+}$ charge. Such radiation-induced loss of H has been already reported 
for mylar \cite{Hac1995}, and for PPS \cite{Kin2021}. 
In contrast to the H yield, the D yield seems to remain unchanged up to 3.0 $\mu$C. 
It is notable that such immobile behavior of deuterium under irradiations of ions of MeV 
energies has been reported for deuterated polystyrene films \cite{Abe1995}. 
Irrespective of such different behavior of H and D, the quantity $R$ needed in the present analysis is 
the ratio of D to H yield in the limit of zero incident He$^{2+}$ charge. 
The values of $R$ can be determined from the extrapolated yield onto the vertical axes in Figs. 4 and 5. 
Actually, the dashed lines shown in Figs. \ref{fig-4} and \ref{fig-5} 
were used for the extrapolation. For mylar and PPS, the dashed lines for the D yield show   
the average of the measured values plotted in Figs. \ref{fig-4} and \ref{fig-5}, and therefore,  
the statistical errors of $R$ result from the total, rather than individual, values of the D yield.
The values of $R$ thus obtained are listed in Table \ref{table-1}.
\begin{table}
\caption {Measured ratios of D to H yield $R$, thereby determined 
D/H ratios, and corresponding D concentrations of the mylar and PPS samples.  }
 \vskip1.5mm
\begin{center}
 \begin{tabular}{lccccr}
 \hline\hline 
  & & \\[-5pt]
  Sample & $R$ ($10^{-4}$) &  D/H ratio ($10^{-4}$) & D concentration (atoms/cm$^{3}$) \\[1.5mm]
\hline 
   & & \\[-5pt]
  Mylar & 1.04\,$\pm$0.05 & 1.60\,$\pm$\,0.08 & $(5.58\pm0.27) \times 10^{18}$ \\[1.5mm]
  PPS & 0.82\,$\pm$0.05 & 1.23\,$\pm$\,0.08 & $(3.70\pm0.23) \times 10^{18}$ \\[2.0mm]
  \hline \hline
 \end{tabular}
\end{center}
\label{table-1}
\end{table}

\subsection{\normalsize D/H ratio}  
\label{DH-ratio}

The D/H ratios can be determined from 
the measured values of $R$. Actually, we may write
\begin{eqnarray}
({\rm D/H\,\, ratio})=R \times \frac{\sigma_{\rm H}}{\sigma_{\rm D}} \;,
\label{eq-1}
\end{eqnarray}
where $\sigma_{\rm H}$ and $\sigma_{\rm D}$ are the recoil cross sections in the laboratory 
system for H and D, respectively, which are given by integrating the differential recoil cross sections 
${\rm d}\sigma / {\rm d}\Omega$ over the solid acceptance angle of the detector $\Omega$. 

In the present narrow angular range of acceptance from 0 to $\varphi_{0} =3.78^{\circ}$,  
${\rm d}\sigma / {\rm d}\Omega$ for H can be assumed to be constant. Actually, it varies only 2\%  
or less, according to the numerical table given by IBANDL \cite{iba}. Similar behavior of 
${\rm d}\sigma / {\rm d}\Omega$ for D is expected from a few available experimental data \cite{Ish2004}. 
Therefore, the ratio of $\sigma_{\rm H}$ to $\sigma_{\rm D}$ in Eq. (\ref{eq-1}) can be replaced by 
the corresponding ratio of ${\rm d}\sigma / {\rm d}\Omega$ at a representative value of the 
recoil angle less than $3.78^{\circ}$.  
For D, we may use the experimental data by Galonsky and coworkers \cite{Gal1955},  
from which ${\rm d}\sigma / {\rm d}\Omega$ at a recoil angle of 3.2$^{\circ}$ can be obtained by 
coordinate transformation from their deuteron--$^{4}$He differential scattering cross sections 
at 173.6$^{\circ}$ in the center of mass system. 

In the above procedure,  
the He$^{2+}$ energies to be considered are not only 8.02 MeV at the surfaces of the films, 
but also 7.81 and 7.92 MeV at the backsurfaces of the mylar and PPS films, respectively. 
The values of ${\rm d}\sigma / {\rm d}\Omega$ at 3.2$^{\circ}$ obtained for D are 
1.28, 1.19, and 1.11 b/sr 
\footnote{The corresponding values in the center of mass system, given in \cite{Gal1955}, 
are 0.320, 0.296, and 0.276 b/sr, respectively.}
for the He$^{2+}$ energies of 8.02, 7.92, and 7.81 MeV, respectively, while  
the corresponding values for H from IBANDL are 1.87, 1.83, and 1.79 b/sr, respectively.  
Therefore, we obtain $\sigma_{\rm H}/\sigma_{\rm D}=1.46$, 1.54, and 1.63    
for He energies of 8.02, 7.92, and 7.81 MeV, respectively. By taking into account the 
small energy losses of He when passing through the mylar and PPS films, the suitable values of 
$\sigma_{\rm H}/\sigma_{\rm D}$ to be used are $(1.46+1.63)/2=1.54$ and $(1.46+1.54)/2=1.50$ for mylar 
and PPS, respectively. The values of D/H obtained from Eq. (\ref{eq-1}) are summarized 
in Table \ref{table-1}, together with the corresponding D concentrations obtained using known 
concentrations of H in mylar and PPS ($3.485 \times 10^{22}$ and $3.006 \times 10^{22}$ cm$^{-3}$, 
respectively). It is noted that the uncertainty of 2.2\% in the experimental cross sections by 
Galonsky and coworkers is only a minor factor in determining the D/H ratios. 

The determined values of D/H in the mylar and PPS films are slightly deviated from 
the Standard Mean Ocean Water (V-SMOW) value of $(1.5576\pm0.0010)\times 10^{-4}$ \cite{DeW1980}.  
To discuss such deviation from the V-SMOW value, accumulation of T-ERDA data with higher sensitivity 
than in the present case is required. Nevertheless, the present work has confirmed the wide applicability 
of T-ERDA to non-destructive determination of D/H ratios in solids. 

Considering the whole analysis procedure to determine the D/H ratios in Table \ref{table-1}, 
we recognize that the advantageous features of T-ERDA is certainly due to the one-dimensional layout 
of the measurement. This simplifies the analysis procedure, for example, the constant 
${\rm d}\sigma / {\rm d}\Omega$ discussed before, hence allows to determine the D/H values 
without using a reference sample. In the present T-ERDA, the uncertainties of the D concentrations 
noted in Table \ref{table-1} corresponds to the sensitivity to 
the amount of D, i.e., $\sim$$10^{17}$ D/cm$^{3}$. 
Typically, the sensitivity is limited by the background yield under the D peak in the spectrum, which 
stems mainly from nuclear reactions occurring in the sample. 

\subsection{\normalsize ($^{4}$He, p) reaction of $^{32}$S}
\label{S-reaction}

We see in Fig. \ref{fig-2}  that the prominent peak at 4.97 MeV shifts to 4.77 MeV for PPS/Ni(25), 
hence the stopping power of Ni is estimated as $200/5=40$ keV/$\mu$m. This value agrees with the stopping 
power for $\sim$5 MeV protons, hence the peak of interest is due to the protons produced possibly by a nuclear reaction in PPS. 
Indeed, the protons result from the $^{32}$S($^{4}$He, p$_{0})^{35}$Cl reaction with the Q value 
equal to $-1.866$ MeV \cite{Moh2015}, where the subscript of p$_{0}$ represents the p-emission with the nuclear transition 
to the ground level of $^{35}$Cl.  

By considering the energy-loss processes of incident He and p$_{0}$ protons 
in PPS/Ni(20), the conservation of momentum and energy including the Q value concludes that the p$_{0}$ yield appears at 
4.96--5.06 MeV. The energy width of 0.10 MeV originates from the energy difference of He at
the surface and the backsurface of PPS of 1.35 $\mu$m thickness, mentioned in \S\ref{DH-ratio}. 
Similarly, the corresponding energies for protons of 
p$_{1}$ and p$_{2}$ which result from transitions to the first (1.22 MeV) and the second (1.76 MeV) excited levels 
of $^{35}$Cl are 3.58--3.69 and 2.95--2.99 MeV, respectively. These values certainly reproduce 
the observed peak energies. 

It is important to note that even a 50 keV shift of the $^{4}$He energy causes a considerable change 
in the reaction cross section of $^{32}$S($^{4}$He, p$_{0})^{35}$Cl \cite{Sol1996}. Because the energy of He 
in collision with S is in the range of 100 keV (7.92--8.02 MeV) in the PPS film (\S\ref{DH-ratio}),
the observed three peaks cannot be directly related to the reaction cross section as a function of He energy.

\section{ \large Conclusions }\label{conclude}

The D/H ratios of mylar and PPS films have been determined by T-ERDA  
without using a reference sample. 
The D/H ratios obtained are only slightly deviated from 
the Standard Mean Ocean Water (V-SMOW) value (1.55$\cdots\times 10^{-4}$),  
indicating potential sensitivity of the present T-ERDA for analysis of 
the variation of ubiquitous deuterium in hydrogen-containing materials. 
Since D/H ratios should reflect chemically synthesized processes experienced by 
the H-containing materials, T-ERDA might lead to unique characterization 
of not only natural, but also synthesized H-containing materials. As a D-sensitive analysis technique, 
T-ERDA will be also useful for development of materials with a controlled 
amount of deuterium. For analysis of spatial distribution of D in a foil sample, 
we may also make use of three-dimensional T-ERDA \cite{Yam2019}, if necessary.  

\newpage
\hspace{-7.5mm} {\bf ACKNOWLEDGEMENTS}\\
\indent The authors thank the technical staff of UTTAC for operating the 
6MV tandem accelerator and Toray KP Films Inc. for providing the PPS films. 
This work was supported in part by JSPS KAKENHI (Grant Number 23K17873) for 
K. Sasa and S. Tomita. 

\newpage
\newpage
\newpage
\newpage
\newpage

\end{document}